\documentclass[showpacs,dvipdfm,aps,amsmath,amssymb,superscriptaddress,notitlepage,twocolumn]{revtex4-1}
\usepackage[dvips]{graphicx}
\usepackage[T1]{fontenc}
\usepackage[utf8]{inputenc}
\usepackage{bm}
\usepackage{pslatex}
\usepackage[]{subfigure}
\usepackage{multirow}
\newcommand{\fref}[1]{Fig.~\ref{#1}}
\renewcommand{\eqref}[1]{Eq.~(\ref{#1})}
\newcommand{\eref}[1]{~(\ref{#1})}

\newcommand{\degree}{\ensuremath{^\circ}}
\usepackage[sort&compress]{natbib}
\usepackage[dvipdfm,pdfauthor={Mirta Herak},pdftitle={Magnetically ordered state of quasi-1D CuSb2O6},colorlinks=true,citecolor=blue,linkcolor=blue,urlcolor=blue,bookmarks=true]{hyperref}
\begin{document}

\title[Magnetically ordered state of quasi-1D CuSb$_2$O$_6$]{\texorpdfstring{Torque magnetometry study of magnetically ordered state and spin reorientation in quasi-1D $\mathbf{S=1/2}$ Heisenberg antiferromagnet CuSb$_\mathbf{2}$O$_\mathbf{6}$}{Torque magnetometry study of magnetically ordered state and spin reorientation in quasi-1D S=1/2 Heisenberg antiferromagnet CuSb2O6}}

\author{Mirta Herak}\email{mirta@ifs.hr}
\affiliation{Institute of Physics, Bijeni\v{c}ka cesta 46, HR-10000 Zagreb, Croatia}
\author{Dijana \v{Z}ili\'{c}}
\affiliation{Ru{\dj}er Bo\v{s}kovi\'{c} Institute, Laboratory for Magnetic Resonances, Bijeni\v{c}ka cesta 54, HR-10000 Zagreb, Croatia}
\author{Dubravka Matkovi\'{c} \v{C}alogovi\'{c}}
\affiliation{University of Zagreb, Faculty of Science, Department of Chemistry, Horvatovac 102A, HR-10000 Zagreb, Croatia}
\author{Helmuth~Berger}
\affiliation{Institute de Physique de la Mati\`{e}re Complexe, EPFL, CH-1015 Lausanne, Switzerland}
\date{\today}
\begin{abstract}
Antiferromagnetically ordered state of monoclinic quasi-one-dimensional $S=1/2$ Heisenberg antiferromagnet CuSb$_2$O$_6$ was studied combining torque magnetometry with phenomenological approach to magnetic anisotropy. This system is known to have a number of different twins in monoclinic $\beta$ phase which differ in orientation of the two CuO$_6$ octahedra in unit cell resulting in different orientation of magnetic axes with respect to crystal axes for each twin.  
We performed torque measurements in magnetic fields $H\leq 0.8$~T on a sample where certain type of twin was shown to be dominant by ESR spectroscopy. The measured data reveal that easy axis is the crystallographic $b$ axis for this sample. Phenomenological magnetocrystalline anisotropy energy invariant to crystal symmetry operations was used to model spin axis direction in zero and finite magnetic field. Our model reproduces the value of the spin flop field $H_{SF}=1.25$~T found in literature. Combination of this approach with our torque results shows that the spin axis will flop in the direction of the maximal value of measured $\mathbf{g}$ tensor when magnetic field $H>H_{SF}$ is applied along easy axis direction. Our analysis of magnetocrystalline anisotropy energy predicts two possibilities for the easy axis direction in this system, $b$ or $a$, connected to different crystallographic twins that can be realized in CuSb$_2$O$_6$. These results offer possibility to reconcile different reports of easy axis direction found in literature for this system and also nicely demonstrate how combination of torque magnetometry and phenomenological approach to magnetic anisotropy can be used to determine the value of the spin flop field and the direction of spin axis in antiferromagnets in both $H<H_{SF}$ and $H>H_{SF}$ by performing measurements in fields significantly smaller than $H_{SF}$. 
\end{abstract}

\pacs{75.25.-j, 75.30.Gw, 75.50.Ee}
\maketitle
\section{Introduction}\label{sec:intro}
Copper oxides represent one of the most studied materials in condensed matter physics. While this has its roots in realization of first high-temperature superconductors in cuprates, nowadays these materials present an arena for study of many interesting and potentially applicable phenomena related to magnetism. Magnetism in copper oxides comes from copper ion Cu$^{2+}$ which has unpaired electron carrying spin $S=1/2$. Magnetic Cu$^{2+}$ ions are treated as spin-only and the effect of their unquenched orbital moment is included in anisotropic electron $\mathbf{g}$ tensor. Distorted and usually elongated CuO$_6$ octahedra found in most copper oxides are responsible for orbital ordering of the $3d$ orbital of the unpaired electron which then influences the dimensionality of the specific magnetic lattice defined by the dominant interactions between spins $S=1/2$ and usually described by the isotropic Heisenberg Hamiltonian $\mathcal{H} = J \sum_{ij} \mathbf{S}_i \cdot \mathbf{S}_j$. In magnetically ordered states, however, magnetic moments choose specific directions in space which are not anticipated by this isotropic model. Weak anisotropic terms must then be added to the spin Hamiltonian which often significantly complicates the theoretical treatment of the properties of these systems. Exact experimental determination of magnetically ordered ground state can be helpful in identifying the relevant anisotropic contributions in specific system.\\ 
\indent In this work we study the antiferromagnetically ordered state of monoclinic quasi-one-dimensional CuSb$_2$O$_6$ motivated by the different reports on easy axis direction and ground state magnetic structure found for this system in literature. We employ torque magnetometry as a sensitive probe of macroscopic magnetic anisotropy. Main advantage of torque magnetometry over magnetic susceptibility measurements is that magnetic torque measures in-plane magnetic susceptibility anisotropy rather than susceptibility along specific axis. This allows not only determination of the direction of easy axis in collinear antiferromagnet, but also predictions about the directions in which the spins will flop when magnetic field along easy axis direction is applied. Our analysis of the interplay of crystal structure and magnetic anisotropy of different crystallographic twins realized in CuSb$_2$O$_6$ in combination with our torque measurements puts forward a possibility to reconcile different reports of the ground state magnetic structure of this system.\\
\indent CuSb$_2$O$_6$ crystallizes in monoclinically distorted trirutile structure type \cite{Nakua-1991} at room temperature. Below 380~K a second order phase transition takes place from high-temperature tetragonal $\alpha-$CuSb$_2$O$_6$ (space group $P4_2/nmn$) to monoclinic $\beta-$CuSb$_2$O$_6$ (space group $P2_1/n$) \cite{Giere-1997, Prokofiev-2003}. This phase transition lifts the degeneracy of four O2 oxygens which exists in the tetragonal phase. Above 380~K dynamic Jahn-Teller effect is realized in the tetragonal phase where higher vibronic states are occupied leading to fast dynamic exchange between the two different possibilities for elongation resulting in averaged compressed CuO$_6$ octahedra \cite{Giere-1997}. Below the phase transition static Jahn-Teller effect is realized through localization of vibronic states in two energy minima of the ground-state potential surface, corresponding to CuO$_6$ octahedra elongated along the Cu-O2a or the Cu-O2 bond \cite{Giere-1997}. The phase transition from tetragonal to monoclinic phase is accompanied by a formation of crystallographic twins which is a result of two different choices for elongation of two crystallographically equivalent CuO$_6$ octahedra in the unit cell of CuSb$_2$O$_6$ \cite{Prokofiev-2003,Heinrich-2003}.\\
\indent Magnetic susceptibility measurements revealed that CuSb$_2$O$_6$ is quasi-one-dimensional antiferromagnet with intrachain exchange interaction $J/k_B\approx 90$~K \cite{Nakua-1991}, which was unexpected given the almost square planar arrangement of magnetic Cu$^{2+}$ ions in the $ab$ plane. Octahedral oxygen environment of the magnetic Cu$^{2+}$ ions gives rise to dominant Cu-O-O-Cu superexchange pathway which forms magnetic chains along the $(a+b)$ direction at $z=0$ and $(a-b)$ direction at $z=0.5$ \cite{Nakua-1995, Koo-2001}. Band structure calculations interpreted the origin of one-dimensional magnetism in CuSb$_2$O$_6$  as being driven by unusual orbital ordering which allows superexchange to be realized through $d_{3z^2-r^2}$ orbitals and apical instead of square planar oxygens \cite{Kasinathan-2008}. These calculations, however, were performed for tetragonal phase of CuSb$_2$O$_6$ where CuO$_6$ octahedra are compressed. In monoclinic phase standard orbital ordering of $d_{x^2-y^2}$  orbitals is expected due to the above mentioned elongation of octahedra \cite{Koo-2001}.\\
\indent Despite the one-dimensional character of the magnetic lattice of CuSb$_2$O$_6$, phase transition to long range antiferromagnetic order takes place below $T_N\approx 8.5$~K due to weak interchain interaction \cite{Nakua-1995, Kato-2002, Gibson-2004,Heinrich-2003}. The reports on magnetic structure and magnetic susceptibility anisotropy are, however, somewhat controversial. Neutron diffraction measurements determined the propagation vector $k=(1/2,0,1/2)$ and reduced magnetic moment of $\mu_{eff}\approx 0.5\mu_B$ \cite{Nakua-1995, Kato-2002,Gibson-2004} and $\mu_{eff}\approx 0.4\mu_B$ in one study \cite{Wheeler-2007}. Magnetic moments within the $ab$ plane are oriented antiparallel along $a$ and $c$ directions and parallel along $b$ direction. Different orientations of magnetic moments with respect to crystal axes are, however, reported in literature. Powder neutron diffraction measurements  revealed spins to be confined to $ab$ plane, but could not distinguish between structure where spins at $z=0$ and $z=0.5$ are collinear and structure where they are canted \cite{Nakua-1995}. Single crystal neutron diffraction studies in one case proposed magnetic structure with moments canted away from the $b$ axis toward the $a$ axis with different canting in $z=0$ and $z=0.5$ layers \cite{Gibson-2004}. This choice of canting was based on the magnetic susceptibility anisotropy reported in the same paper \cite{Gibson-2004}. Two other single crystal neutron diffraction studies showed that collinear antiferromagnetic order is realized with spins aligned in the $b$ axis direction, although some weak reflections could not be indexed \cite{Kato-2002}, or were disregarded as signal from crystal twins \cite{Wheeler-2007}.\\
\indent Below $T_N$ different magnetic susceptibility anisotropy reports are found in literature. Large decrease of susceptibility characteristically observed when magnetic field is applied along the easy axis has been reported for $b$ axis direction \cite{Heinrich-2003, Torgashev-2003}, $a$ axis direction \cite{Prokofiev-2003}, and even for both directions \cite{Gibson-2004, Rebello-2013,Christian-2014}. Recent report on single crystal measurements of magnetic susceptibility showed that significant decrease of susceptibility below $T_N$ and spin flop transition at $H_{SF}=1.25$~T is observed when magnetic field is applied along both the $a$ and $b$ axis, suggesting that both $a$ and $b$ are the easy axes in CuSb$_2$O$_6$ \cite{Rebello-2013}. The magnetically ordered ground state of CuSb$_2$O$_6$ is thus still not completely resolved.\\
\indent In this work we attempt to shed some light on magnetically ordered ground state of CuSb$_2$O$_6$ using torque magnetometry in magnetic field smaller than the spin-flop field. We combine our torque measurements with phenomenological approach to magnetic anisotropy where we model the magnetocrystalline anisotropy using crystal and magnetic symmetry restrictions to describe the anisotropic magnetic response of this system in finite magnetic field. This approach allows us to determine the orientation of spin axis in both zero field and in $H>H_{SF}$. In Sec. \ref{sec:sample} we describe the characterization of our sample by X-ray diffraction, magnetic susceptibility measurements and electron spin resonance (ESR) spectroscopy. In Sec. \ref{sec:torqueAFM} the results of torque magnetometry are presented. Phenomenological approach to magnetic anisotropy of CuSb$_2$O$_6$ is used for the description of measured torque curves in Sec. \ref{sec:Faniso}. The results are discussed and the conclusion given in Secs. \ref{sec:disc} and \ref{sec:concl}, respectively.
\section{Sample characterization}\label{sec:sample}
Single crystals of CuSb$_2$O$_6$ were grown using endothermic chemical vapor transport reactions with different transport agents. Firstly, polycrystalline CuSb$_2$O$_6$ powder was synthesized by solid state reaction described in detail elsewhere \cite{Prokofiev-2003}. Secondly, CuSb$_2$O$_6$ powders and HCl or TeCl$_4$ were mixed and sealed in a 24~mm diameter and 20~cm long evacuated quartz tube. The tube was inserted in a two-zone horizontal tube furnace with a source-zone temperature of 900$\degree$C and a growth-zone temperature of 800$\degree$C for four weeks. In this way, millimeter-sized single crystals of CuSb$_2$O$_6$ were obtained. No impurity phases were detected using X-ray powder diffraction. The crystals are yellow in color, as reported in literature \cite{Prokofiev-2003}.\\
\indent One large crystalline sample was ground into powder and used for measurements of average magnetic susceptibility. Small thin crystal with mass $m=(91\pm 2)~\mu$g cut from larger crystal was used for torque measurements. On the surface of the sample thin stripes due to twinning were visible \cite{Prokofiev-2003}. We did not attempt to detwin the sample. This sample was characterized by X-ray diffraction and ESR spectroscopy.\\

\subsection{X-ray diffraction}\label{sec:xray}
The orientation of the crystal axes with respect to the sample morphology was determined using X-ray diffraction. The crystal was glued onto a glass fiber and the data were collected on an Oxford Diffraction Xcalibur 3 CCD diffractometer with graphite-monochromated Mo $K_{\alpha}$ radiation ($\lambda = 0.71073$~\AA) at room temperature. The orientation of the crystal and the unit cell parameters were obtained by the CrysAlis software package by Agilent Technologies. Data confirmed the monoclinic space group $P2_1/n$ with the following unit cell parameters: $a = 4.6299(3)$~\AA, $b = 4.6343(3)$~\AA, $c = 9.2911(6)$~\AA, $\beta = 90.965(7)\degree$, in good agreement with previously published data \cite{Nakua-1991}. Presence of multiple twins, reported by other authors \cite{Prokofiev-2003, Giere-1997,Heinrich-2003,Rebello-2013,Christian-2014,Prasai-2015}, was also observed in our data. \\
\subsection{Magnetic susceptibility}\label{sec:susc}
\begin{figure}[tb]
	\centering
		\includegraphics[width=1.00\columnwidth]{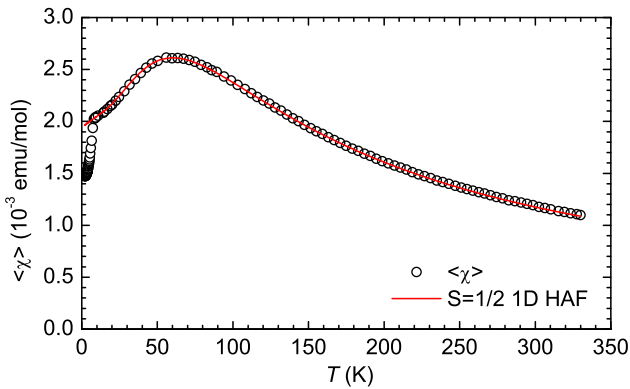}
	\caption{Magnetic susceptibility of powder CuSb$_2$O$_6$. Solid line represents fit to magnetic susceptibility of spin $S=1/2$ 1D Heisenberg antiferromagnet \cite{Bonner-Fisher,Johnston-2000}.}
	\label{fig1}
\end{figure}

Magnetic susceptibility of powder, $\left< \chi \right>$, was measured using Faraday apparatus in the temperature range from 2~K to 330~K in magnetic field $H=7$~kOe. Temperature dependence of magnetic susceptibility $\left< \chi \right>$ of CuSb$_2$O$_6$ measured on powder is shown in \fref{fig1}. The measured data can be described by the susceptibility of one-dimensional (1D) spin $S=1/2$ Heisenberg antiferromagnet with $J/k_B=94$~K and $\left<g \right>=2.11$ \cite{Bonner-Fisher,Johnston-2000}, in agreement with results reported in literature \cite{Nakua-1991,Kato-2002,Prokofiev-2003,Heinrich-2003,Gibson-2004,Rebello-2013}. At $T_N\approx 8.5$~K a kink in susceptibility marks the phase transition to long range magnetic order. Below $T_N$ $\left< \chi \right>$ decreases as temperature decreases in a manner typical for collinear antiferromagnet \cite{Kittel}. To determine the direction of the easy axis we employ torque magnetometry in Section \ref{sec:torqueAFM}.
\subsection{ESR spectroscopy}\label{sec:ESR}
\begin{figure}[t]
	\centering
		\includegraphics[width=1.00\columnwidth]{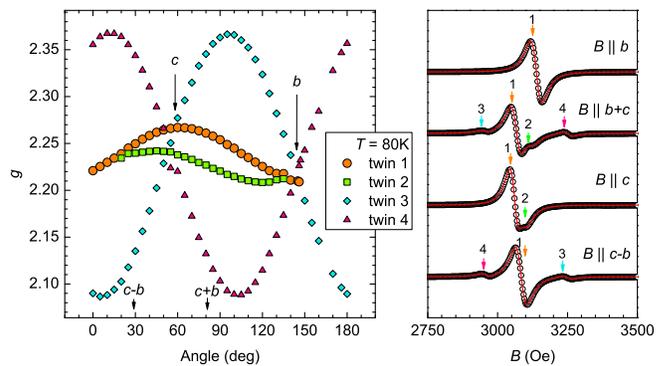}
	\caption{X-band ESR spectra (right) and $g$ factor anisotropy (left) measured in the $bc$ plane at $T=80$~K. Solid lines in figure on the right represent results of fitting of spectra to four individual Lorentzian lineshapes. Error bars for $g$ factors obtained from the fit are smaller than the size of symbols.}
	\label{fig2}
\end{figure}

\begin{figure}[t]
	\centering
		\includegraphics[width=1\columnwidth]{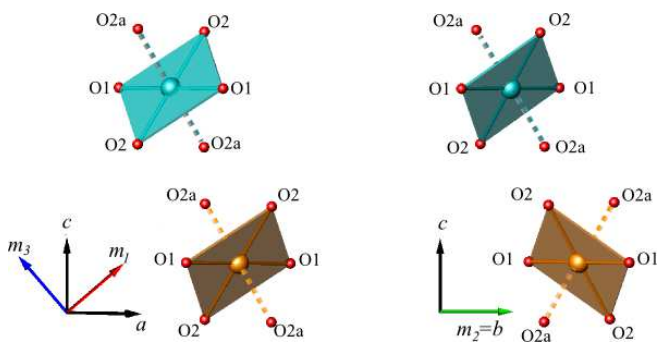}
	\caption{Two magnetically inequivalent CuO$_6$ octahedra in twin I of $\beta-$CuSb$_2$O$6$ as viewed in the $ac$ and the $bc$ plane. Lower (orange) is represented as Oct1 in table \ref{tab1}, and upper (blue) as Oct2. Crystal axes directions are shown in black, and macroscopic magnetic axes $m_1$, $m_2$ and $m_3$ for twin I are shown in red, green and blue, respectively.}
	\label{fig3}
\end{figure}

X-band ESR measurements were performed at $T=80$~K using Bruker Elexsys 580 FT/CW X-band spectrometer equipped with a standard Oxford Instruments model DTC2 temperature controller. The microwave frequency was $\sim 9.7$~GHz with the magnetic field modulation amplitude of 5~G at 100~kHz. ESR spectra were recorded at $5\degree$ steps and the rotation was controlled by a home-built goniometer with the accuracy of $1\degree$. The uncertainty of $\approx 5\degree$ was related to the optimal deposition of the crystal on the quartz sample holder.\\
\indent We performed ESR spectroscopy measurements at $T=80$~K to determine the number and type of twins in our single crystal sample which is used for torque magnetometry study described in Sec. \ref{sec:torqueAFM}. Previous detailed ESR analysis of CuSb$_2$O$_6$ has shown that there are four possible twin domains in the monoclinic $\beta$ phase which appear simultaneously during the structural phase transition \cite{Giere-1997,Heinrich-2003}.\\
\indent Spectra measured at several different directions in the $bc$ plane of our sample are shown in \fref{fig2}. Four ESR peaks of different intensity could be detected while for $H||b$ only one single Lorentzian line is observed. Two central lines, one of which was the most intensive line in the spectra (twin 1) and the other was weak (twin 2), are located at approximately the same resonant fields and therefore it was difficult to distinguish them. Each of the four lines can be explained by the anisotropic Zeeman interaction due to the anisotropic $g$ tensors of the two magnetically inequivalent Cu$^{2+}$ ions in the unit cell which merge to one resonance due to strong exchange coupling \cite{Heinrich-2003}. To obtain angular dependence of $g$ factor for each line, when possible, we fitted our spectra to four individual Lorentzian lines parametrized by its resonance field $H_{res}\propto g$, linewidth $\Delta H$ and intensity. The results of fit are shown as solid lines describing the spectra in \fref{fig2}. Due to low resolution of our X-band spectra, the contributions of twins 1 and 2 were difficult to distinguish at almost all angles, and contributions of all four twins were difficult to distinguish around $b$ and $c$ axes where spectra of different twins merge. Intensities resulting from the fit suggest that twins 3 and 4 make up for less than 10\% of our sample. \\
{\renewcommand{\arraystretch}{1.5}
\begin{table*}[bt]
	\centering
	\begin{ruledtabular}
		\begin{tabular*}{\linewidth}{@{\extracolsep{\fill}}c| c c c c c c c c}
			\multirow{2}{*}{Twin} & \multirow{2}{*}{Symmetry} & Elongated& Elongated & \multirow{2}{*}{$m_1$} & \multirow{2}{*}{$m_2$} & \multirow{2}{*}{$m_3$} & easy  & \multirow{2}{*}{$H_{||ea}>H_{SF}$} \\
			& &  Cu-O in Oct1 & Cu-O in Oct2 & & & & axis &\\
			\hline \hline
			\textrm{I} & \multirow{2}{*}{$P 1 2_1/n 1 $} & Cu-O2a & Cu-O2a &  $(\gamma, 0, \eta)$ & $\hat{b}$ & $(-\eta, 0,\gamma)$ & $\hat{b}$ & $m_3$\\
			\textrm{II} &  & Cu-O2& Cu-O2 & $(\gamma, 0, -\eta)$ & $\hat{b}$ & $(\eta, 0, \gamma)$ & $\hat{b}$ & $m_3$\\
			\textrm{III} & \multirow{2}{*}{$P 2_1/n 1 1 $\footnote{This choice of space group and consequently direction of easy axis is obtained if common crystal axes are assumed for all twins, as explained in the text.}} & Cu-O2a & Cu-O2& $(0, -\gamma, \eta)$ & $\hat{a}$ & $(0, \eta, \gamma)$ & $\hat{a}$ & $m_3$\\
			\textrm{IV}  & & Cu-O2& Cu-O2a & $(0, -\gamma, -\eta)$ & $\hat{a}$ & $(0, -\eta, \gamma)$ & $\hat{a}$ & $m_3$
		\end{tabular*}
		\end{ruledtabular}
	\caption{Elongated Cu-O bonds and directions of macroscopic magnetic axes $m_1$, $m_2$ and $m_3$ with respect to the crystal axes for each twin. $m_1$, $m_2$ and $m_3$ are unit vectors. Oct1 (Oct2) is lower orange (upper blue) octahedron in \fref{fig3}. Direction $m_3$ is determined as the direction of the maximal value of the $g$ tensor for each twin. $\gamma=\cos(\sphericalangle(m_3, \hat{c}))$ and $\eta=\sqrt{1-\gamma^2}$. Easy axis direction for each twin is given in column before last. The direction of spin axis when field $H_{||ea}>H_{SF}$ is applied along the easy axis direction is $m_3$ in our model, where $m_3$ axis for each twin is given in the second column before last.}
	\label{tab1}
\end{table*}
}
\indent Appearance of the observed four different twins in $\beta$-CuSb$_2$O$_6$ has its origin in a change occurring in CuO$_6$ octahedra as they go through phase transition from high-temperature tetragonal to the low-temperature monoclinic phase. There are two differently oriented but crystallographically equivalent CuO$_6$ octahedra in the unit cell of CuSb$_2$O$_6$. In the tetragonal $\alpha$ phase CuO$_6$ octahedra are compressed with the direction of compression being along the Cu-O1 bonds, and four equivalent Cu-O2 bonds in the equatorial plane. In the monoclinic $\beta$ phase CuO$_6$ octahedra become elongated along two of the Cu-O2 bonds which now become Cu-O2a bonds, while the other two Cu-O2 bonds shrink to form an equatorial plane with Cu-O1. Direction of compression in $\alpha$ phase and elongation in $\beta$ phase is thus not the same. Four different twins can emerge from two different choices for elongation for each of the two magnetically inequivalent CuO$_6$, as already discussed in Ref. \onlinecite{Giere-1997} and evidenced in ESR measurements \cite{Heinrich-2003}. Using terminology of Ref. \onlinecite{Heinrich-2003}, we label those twins in roman letters from \textrm{I} to \textrm{IV}. In \fref{fig3} we show the two CuO$_6$ octahedra for twin \textrm{I}, Oct1 and Oct2, in orange and blue. In table \ref{tab1} we summarize the elongated Cu-O bonds for four possible crystallographic twins.\\
\indent From the above presented analysis and the results of Ref. \onlinecite{Heinrich-2003} it follows that lines of twins \textrm{I} and \textrm{II} should be indistinguishable in the $bc$ plane. The splitting we observe for our twins 1 and 2 could result from the misorientation of the sample. The values of $g$ factor we obtain for twin 1 in the $bc$ plane are $g_b=2.21$ and $g_c=2.27$. For twin 2 we observe $g_c=2.24$. This compares rather well to the data from literature, $g_b=2.21$, $g_c=2.23$ given in Ref. \onlinecite{Heinrich-2003}. Our twins 3 and 4 correspond to twins \textrm{III} and \textrm{IV} in Ref. \onlinecite{Heinrich-2003}, respectively. \\
\indent Local magnetic axes, $x,\; y$, and $ z$ of each CuO$_6$ octahedron are defined by the CuO$_4$ plaquette of four nearest oxygens \cite{Giere-1997,Heinrich-2003}. Local $x$ axis is in the direction of Cu-O1, $y$ is in the plaquette, nearly in the Cu-O2 direction, and $z$ is perpendicular to the plaquette \cite{Giere-1997,Heinrich-2003}. $z$ axis is thus $\approx 12\degree $ away from the apical Cu-O2a bond of the octahedron. This choice of local magnetic axes was corroborated by the measured $g$ factor anisotropy for each twin \cite{Heinrich-2003}. The resultant macroscopic magnetic axes of the sample, $m_1$, $m_2$ and $m_3$, are thus not in the direction of the crystal axes. Direction of macroscopic magnetic axes in Cu$^{2+}$ $S=1/2$ systems can be determined experimentally as the axes in which the $\mathbf{g}^2$ tensor and magnetic susceptibility tensor $\bm{\hat{\chi}}$ are diagonal.\\ 
\indent Using the above given local magnetic axes we obtain for twin \textrm{I} that macroscopic magnetic axis $m_2$ is in the direction of the $b$ axis, $m_2 = \hat{b}$, and $m_1$ and $m_3$ are in the $ac$ plane. The angle between axes $m_1$ and $c$ in twins \textrm{I} and \textrm{II} amounts to $\sphericalangle(m_1, \hat{c})\approx 49\degree$, and $\sphericalangle(m_3, \hat{c}) \approx 41\degree$, but the orientation of these axes is different in those twins, as evidenced from ESR measurements \cite{Heinrich-2003}. In twins \textrm{III} and \textrm{IV} $b$ axis is no longer magnetic axis, as was observed in previously reported ESR measurements \cite{Heinrich-2003}, and as can be seen from our results for twins 3 and 4 shown in \fref{fig2}. This has consequences for the crystal symmetry, as we will now argue. In twins \textrm{I} and \textrm{II} $b$ axis is one of the magnetic axes and the other two magnetic axes are in the $ac$ plane and are rotated from $a$ and $c$ crystal axes, as evidenced from ESR measurements \cite{Heinrich-2003}. According to Neumann's principle any physical property of a crystal must be invariant to the same point group symmetry operations as the crystal structure \cite{Newnham}. For CuSb$_2$O$_6$ this means that measured $\mathbf{g}^2$ tensor and susceptibility tensor must be invariant to $2/m$ symmetry operations \footnote{$2/m$ symmetry operations are rotation of 180 \degree around unique axis and mirror plane perpendicular to unique axis.} with respect to $b$ axis for conventional choice of $b$ axis as unique axis in $P 2_1/n$ space group, which can also be written as $P 1 2_1/n 1$. For twins \textrm{I} and \textrm{II} that is indeed the case. For twins \textrm{III} and \textrm{IV} $b$ axis is no longer magnetic axis, as can be seen from \fref{fig2} and also in Ref. \onlinecite{Heinrich-2003}. Magnetic axes for those twins are axis $m_2=\hat{a}$, and axes $m_1$ and $m_3$ which correspond to directions of minimal and maximal value of $\mathbf{g}$ in the $bc$ plane. Magnetic properties such as $\mathbf{g}^2$ tensor and magnetic susceptibility tensor are now invariant to $2/m$ symmetry elements with respect to the $a$ axis. According to Neumann's principle this means that unique axis is $a$ axis for twins \textrm{III} and \textrm{IV} and their space group should be labeled as $P 2_1/n 1 1 $. We distinguish these two symmetry variations of $P2_1/n$ in table \ref{tab1}. This will be helpful in construction of magnetocrystalline anisotropy energy for each of the twins in Sec. \ref{sec:Faniso}.\\  
\indent Another possibility of interpreting $g$ factor data for twins \textrm{III} and \textrm{IV} would be by assuming that these twins have axes $a$ and $b$ rotated by 90$\degree$ with respect to those of twins \textrm{I} and \textrm{II}. We are not able to either support or dispute this possibility from our structural measurements and since it was not suggested by previous structural measurements \cite{Giere-1997,Prokofiev-2003,Heinrich-2003} we continue by assuming common crystal axes for all twins in our sample.
\section{Results}\label{sec:results}
\subsection{Torque magnetometry in the AFM state of $\mathbf{CuSb_2O_6}$}\label{sec:torqueAFM}
Magnetic torque was measured with home-built torque apparatus which uses torsion of thin quartz fiber for torque measurement. The sample holder is made of ultra pure quartz and has an absolute resolution of $10^{-4}$~dyn~cm. The uncertainty of $\approx 5\degree$ was related to the optimal deposition of the crystal on the quartz sample holder. This uncertainty was somewhat larger for the $a^*b$ plane because of sample's morphology.\\
\indent Magnetic torque $\bm{\tau}$ acting on a sample of volume $V$ with magnetization $\mathbf{M}$ in magnetic field $\mathbf{H}$ is given by expression
\begin{equation}\label{eq:torque}
	\bm{\tau} = V \; \mathbf{M} \times \mathbf{H}
\end{equation}
In our experiment magnetic field is rotated in an $xy$ plane and only $\tau_z$ component of torque perpendicular to that plane is measured. In case of linear response of induced magnetization to applied magnetic field, $\mathbf{M} = \bm{\hat{\chi}} \cdot \mathbf{H}$, where $\bm{\hat{\chi}}$ is magnetic susceptibility tensor, we obtain for the measured component of torque
\begin{equation}\label{eq:torquePM}
	\tau_z = \dfrac{m}{2M_{mol}} \: H^2 \: (\chi_x -\chi_y) \sin 2\theta.
\end{equation}
In expression \eref{eq:torquePM} $m$ is the mass of the sample, $M_{mol}$ is molar mass, $\chi_x$ and $\chi_y$ are susceptibility tensor components along directions $x$ and $y$ expressed in emu/mol and $\theta$ is the angle magnetic field makes with the $x$ axis. Directions $x$ and $y$ represent directions along which the torque is zero, i.e. magnetization is parallel to applied magnetic field. These directions are the directions of magnetic axes in the $xy$ plane. From expression \eref{eq:torquePM} we see that measured component of torque is proportional to magnetic susceptibility anisotropy, $\Delta \chi_{xy}=\chi_x -\chi_y$, in the plane of rotation of magnetic field. This result is applicable to paramagnets and also collinear antiferromagnets when applied magnetic field is much smaller than the spin flop field $H_{SF}$. Measurements in the paramagnetic state (not shown here) were performed on a larger sample. Angular dependence obeys \eqref{eq:torquePM} and magnetic susceptibility anisotropy displays the same temperature dependence as susceptibility with wide maximum around $\approx 60$~K.\\
\begin{figure}[tb]
	\centering
		\includegraphics[width=1.00\columnwidth]{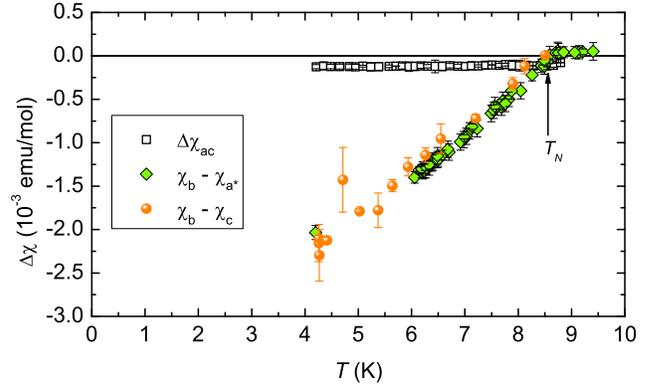}
	\caption{Temperature dependence of magnetic susceptibility anisotropy measured in $ac$, $a^*b$ and $bc$ planes. }
	\label{fig4}
\end{figure}
\begin{figure}[tb]
	\centering
		\includegraphics[width=1\columnwidth]{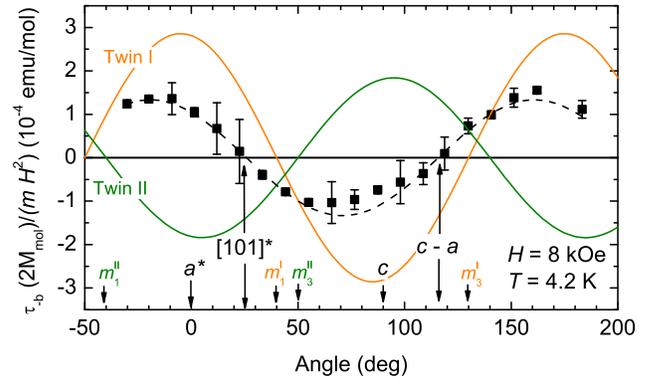}
	\caption{Angular dependence of torque $\tau$ multiplied by $2 M_{mol}/(m H^2)$ [see \eqref{eq:torquePM}] measured in $H=8$~kOe in the $ac$ plane. Macroscopic magnetic axes $m_1^{I}$ and $m_3^{I}$ for twin \textrm{I} and $m_1^{II}$ and $m_3^{II}$ for twin \textrm{II} are also shown. Solid lines represent contributions from twins \textrm{I} and \textrm{II} resulting in dashed line.}
	\label{fig5}
\end{figure}
\indent Temperature dependence of magnetic susceptibility anisotropy measured in magnetically ordered state below $T_N\approx 8.5$~K is shown in \fref{fig4}. Angular dependence of torque multiplied by $2 M_{mol}/(m H^2)$ [see \eqref{eq:torquePM}] measured at $T=4.2$~K in the $ac$, $a^*b$ and $bc$ plane is shown in Figs. \ref{fig5}, \ref{fig6} and \ref{fig7}, respectively. Maxima i.e. minima of shown torque curves represent values of susceptibility anisotropy in those planes. Magnetic susceptibility anisotropy shown in \fref{fig4} was measured by applying magnetic field in the direction where angular dependencies shown in Figs. \ref{fig5}, \ref{fig6} and \ref{fig7} have minima. Below $T_N=8.5$~K large susceptibility anisotropy develops in the planes containing the $b$ axis, $a^*b$ and $bc$ plane, while in the $ac$ plane susceptibility anisotropy remains comparable to its value above $T_N$. Susceptibility anisotropies $\Delta \chi_{ba^*}=\chi_b-\chi_{a^*}$ and $\Delta \chi_{bc}=\chi_b-\chi_c$ amount to $-2.2\cdot 10^{-3}$~emu/mol at $T=4.2$~K, which is slightly larger than the value of $\left<\chi \right>$ at $T_N$, see \fref{fig1}. In collinear antiferromagnet susceptibility anisotropy at $T=0$ amounts to $\chi_{\perp}-\chi_{\|}\approx \chi_{\perp}$, where $\chi_{\perp}$ is expected to be temperature independent below $T_N$ \cite{Kittel}. Since $T_N=8.5$~K in CuSb$_2$O$_6$, measured susceptibility anisotropies in planes with $b$ axis will become even larger when $T<4.2$~K. Observed magnitude of susceptibility anisotropy in comparison with $\left<\chi \right>$ at $T_N$ siginifies that in our sample (or at least the great majority of it) $b$ axis is the easy axis.\\
\indent Angular dependence of torque multiplied by $2 M_{mol}/(m H^2)$ measured at $T=4.2$~K in the $ac$ plane shown in \fref{fig5} can be described by \eqref{eq:torquePM}, although the signal is very weak even in the largest applied field of $H=8$~kOe. If we fit the data to \eqref{eq:torquePM} we obtain the zeros of the sine curves in directions $\approx [101]^*$ and $c-a$, with $\chi_{c-a}>\chi_{[101]^*}$. In \fref{fig5} positions of macroscopic magnetic axes, $m_1$ and $m_3$ for twin \textrm{I} and \textrm{II} are also marked. Since our ESR results suggest that twins \textrm{I} and \textrm{II} make up for majority of the sample, we fitted the observed torque in the $ac$ plane to $\tau=A_I \sin(2\theta-2\theta_{m_3^{I}})+A_{II} \sin(2\theta-2\theta_{m_3^{II}})$, where $A_I$ and $A_{II}$ are amplitudes and $\theta_{m_3^{I}}$ and $\theta_{m_3^{II}}$ angles of axes $m_3^{I}$ and $m_3^{II}$ of twins \textrm{I} and \textrm{II}, respectively. The results of the contributions for each twin are shown in \fref{fig5} by solid lines, and sum of those contributions by dashed line.\\
\begin{figure}[t]
	\centering
		\includegraphics[width=1\columnwidth]{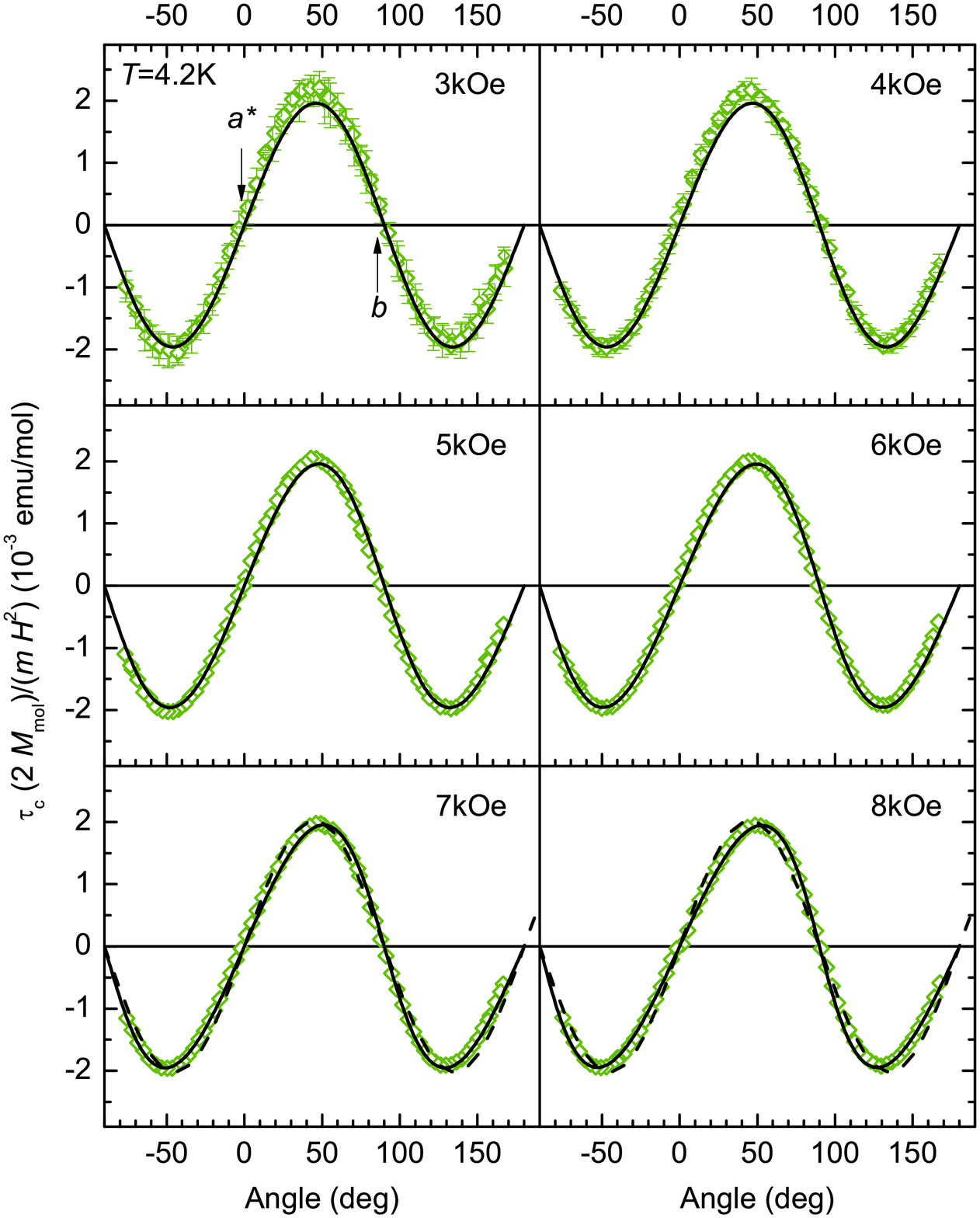}
	\caption{Angular dependence of torque multiplied by $2 M_{mol}/(m H^2)$ [see \eqref{eq:torquePM}] measured in different fields in the $a^*b$ plane. Dashed line represents \eqref{eq:torquePM} expected when no spin reorientation takes place. Solid lines represent results of simulation discussed in text. Error bars are shown only when they are larger than the symbol.}
	\label{fig6}
\end{figure}
\begin{figure}[t]
	\centering
		\includegraphics[width=1\columnwidth]{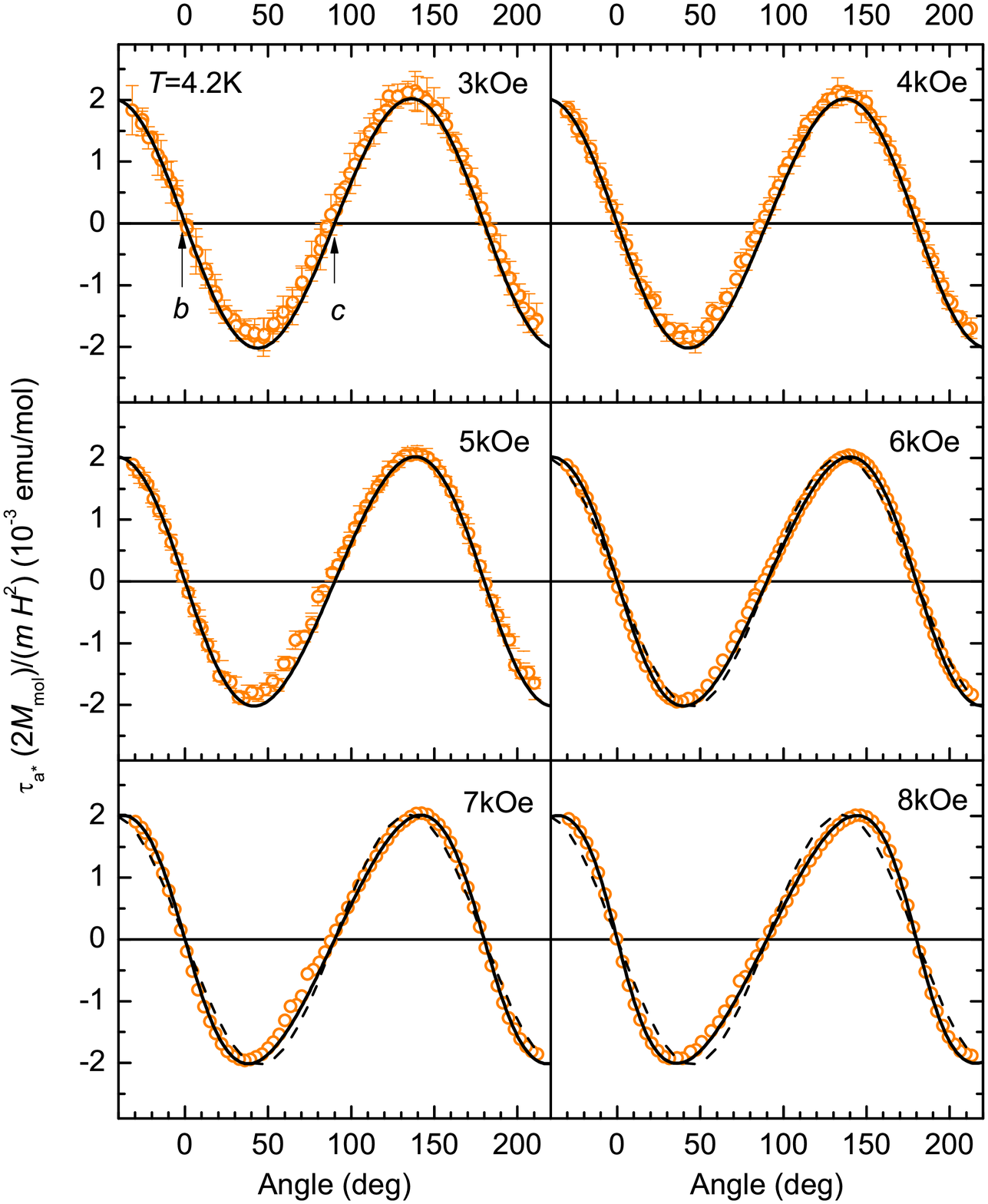}
	\caption{Angular dependence of torque multiplied by $2 M_{mol}/(m H^2)$ [see \eqref{eq:torquePM}] measured in different fields in the $bc$ plane. Dashed line represents \eqref{eq:torquePM} expected when no spin reorientation takes place. Solid lines represent results of simulation discussed in text. Error bars are shown only when they are larger than the symbol.}
	\label{fig7}
\end{figure}
\indent Angular dependence of torque multiplied by $2 M_{mol}/(m H^2)$ in the $a^*b$ and the $bc$ plane measured at $T=4.2$~K, shown in Figs. \ref{fig6} and \ref{fig7} respectively. Amplitude of $\tau \cdot (2M_{mol})/(m H^2)$ does not change with applied field, signifying linear dependence of magnetization on magnetic field. In low magnetic fields, $H\lesssim 5$~kOe, angular dependence in both planes follows \eqref{eq:torquePM}. In higher fields, however, a deviation from this dependence is observed. The deviation is most easily observed as a shift of the angle of maximum and minimum of torque away from the angle it is situated on in lower field. The deviation grows with applied magnetic field and it is slightly larger in the $bc$ plane. The observed deviation of torque is a result of reorientation of spin axis from the easy axis direction and is typical for uniaxial antiferromagnets when applied magnetic field becomes comparable to the spin flop field $H_{SF}$ \cite{Yosida-1951, Uozaki-2000}. In our case the deviation is observed in both $a^*b$ and $bc$ plane, although it is stronger in the latter. In the next section, we study influence of the magnetic anisotropy on the torque curves of this system using phenomenological approach in which magnetic anisotropy is described by the symmetry-obeying magnetocrystalline anisotropy energy. Comparison with measured data will allow us to determine the spin axis direction in both zero field and for $H_{||b}>H_{SF}$.
\subsection{Phenomenological approach to magnetic anisotropy in CuSb$_2$O$_6$}\label{sec:Faniso}
Effect of spin reorientation phenomena in uniaxial antiferromagnets on the torque curves has been successfully studied using phenomenological expression for uniaxial anisotropy energy \cite{Yosida-1951, Uozaki-2000}. This approach has often been limited to only one plane in which the spins are allowed to rotate \cite{Uozaki-2000}. In this work we use the three-dimensional anisotropy energy allowed by symmetry where we allow the spin axis to rotate away from the easy axis in the direction which minimizes the total energy in finite magnetic field. This rotation is described by the rotation of the susceptibility tensor. From the rotated susceptibility tensor we can calculate the magnetization and, using \eqref{eq:torque}, the torque. This approach was used recently to describe the measured torque curves in antiferromagnetically ordered phases of tetragonal Bi$_2$CuO$_4$ in terms of easy plane anisotropy \cite{Herak-2010}, and cubic Cu$_3$TeO$_6$ in terms of cubic anisotropy \cite{ Herak-2011}.\\
\begin{figure}[tb]
	\centering
		\includegraphics[width=0.6\columnwidth]{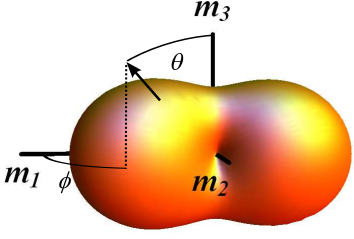}
	\caption{Magnetocrystalline anisotropy energy allowed by symmetry in CuSb$_2$O$_6$, \eqref{eq:Faniso}, with $K'_1>K_1$. Direction of magnetic axis $m_1$, $m_2$ and $m_3$ with respect to crystal axes for each twin is given in table \ref{tab1}.}
	\label{fig8}
\end{figure}
Magnetic susceptibility tensor for system with monoclinic space group $P 1 2_1/n 1$  written in $(a^*, b, c)$ coordinate system is given by \cite{Newnham}
\begin{equation}\label{eq:tensorchiasbc}
	\bm{\hat{\chi}}^{a^{*}bc} = 
	\begin{pmatrix}
	\chi_{a^*} & 0 & \chi_{ac}\\
	0 & \chi_b & 0 \\
	\chi_{ac} & 0 & \chi_c
	\end{pmatrix}.
\end{equation}
\indent The above given expression for susceptibility tensor is valid for twins \textrm{I} and \textrm{II}. Assuming all twins have common $a$ and $b$ axes, for twins \textrm{III} and \textrm{IV} space group $P 2_1/n 1 1$ is more appropriate and unique axis is the $a$ axis, as already discussed in Sec. \ref{sec:ESR}. The coordinate system used in that case is $(a, b^*, c)$ and the only nonzero nondiagonal element of the susceptibility tensor is then $\chi_{bc}$.\\
\indent Magnetic susceptibility tensor is diagonal in coordinate system spanned by  the macroscopic magnetic axes $(m_1, m_2, m_3)$, where $m_1$, $m_2$ and $m_3$ for each twin are given in table \ref{tab1}. In what follows we use coordinate system spanned by magnetic axes $(m_1, m_2, m_3)$. \\
\indent The most general second-order magnetocrystalline anisotropy energy invariant to symmetry operations of monoclinic $\beta$ phase of CuSb$_2$O$_6$ can be written as \cite{Skomski}
\begin{equation}\label{eq:Faniso}
	\mathcal{F}_a  = K_1 \sin^2 \theta + K'_{1} \sin^2 \theta \cos 2\phi,
\end{equation}
where $K_1$ and $K'_1$ are phenomenological anisotropy constants and $\theta$ and $\phi$ are polar and azimuthal angles, respectively. To describe the observed result that axis $m_2$ is the easy axis $(b)$ in CuSb$_2$O$_6$ we set $K'_1>K_1$. This gives the anisotropy energy a global minimum at angles $(\theta,\phi)=(\pi/2,\pi/2)$ to which we assign the direction of $m_2$. The shape of this anisotropy energy is shown in \fref{fig8}. The saddle point represents a direction in which the spins will reorient when magnetic field $H\geq H_{SF}$ is applied in the easy axis direction $m_2$. By symmetry requirements other two axes corresponding to saddle point and global maximum can be directed anywhere in the plane spanned by global magnetic axes $m_1$ and $m_3$. Magnetic symmetry, however, prefers these directions to be in the directions of the magnetic axes. This leaves us with two choices: preferred direction of spins when they flop (the saddle point) is along $m_1$ or $m_3$, the directions of minimal and maximal components of magnetic susceptibility in the $(m_1, m_3)$ plane, respectively. We performed calculations for both and our results favor $m_3$ in the direction of the saddle point, such as shown in \fref{fig8}.\\
\indent Magnetocrystalline anisotropy energy, \eqref{eq:Faniso}, gives us the direction of the easy axis, i.e. spin axis in zero magnetic field. In experiment finite magnetic field $\mathbf{H}(\xi, \psi)=H_0 (\cos \xi \sin \psi, \sin \xi \sin \psi, \cos \psi)$ is applied, where $H_0$ is the magnitude of the field, $\psi$ is the polar angle the field makes with axes $m_3$ and $\xi$ is the azimuthal angle that the projection of $\mathbf{H}$ on the $(m_1,m_2)$ plane makes with the axes $m_1$. The Zeeman term, $\mathcal{F}_Z$, can be written as
\begin{equation}\label{eq:FZeeman}
	\mathcal{F}_Z = - \dfrac{1}{2}\; \mathbf{H} \cdot \bm{\hat{\chi}} \cdot \mathbf{H}.
\end{equation}
In our calculations we use measured components of magnetic susceptibility expressed in emu/mol and magnetic field $H$ expressed in Oe. This gives $\mathcal{F}_Z$ expressed in erg/mol, so our anisotropy constants in expression \eref{eq:Faniso} for anisotropy energy are given in units erg/mol.
In low magnetic field, $H\ll H_{SF}$, the susceptibility tensor in \eqref{eq:FZeeman} is diagonal
\begin{equation}\label{eq:tensorchidia}
	\bm{\hat{\chi}}_{dia} = 
	\begin{pmatrix}
	\chi_{m_1} & 0 & 0\\
	0 &\chi_{m_2} & 0 \\
	0 & 0 & \chi_{m_3}
	\end{pmatrix}.
\end{equation}
As magnetic field increases the spin axis start to reorient from the easy axis direction to minimize the total energy. Following procedure for in-plane reorientation in uniaxial antiferromagnets \cite{Yosida-1951} and more general reorientation in easy plane \cite{Herak-2010} and cubic \cite{Herak-2011} antiferromagnets, we describe the reorientation of spin axis as a rotation of the susceptibility tensor
\begin{equation}\label{eq:tensorchirot}
	\bm{\hat{\chi}}(\theta,\phi) = \mathbf{R}(\theta,\phi) \cdot\bm{\hat{\chi}}_{dia} \cdot \mathbf{R}^T(\theta,\phi),
\end{equation}
where $\mathbf{R}(\theta,\phi)= \mathbf{R}_{m_3}(\phi-\pi/2)\cdot \mathbf{R}_{m_1} (\theta-\pi/2) $ is the matrix that describes the rotation of the susceptibility tensor \footnote{$\mathbf{R}_i(\varphi)$ is the rotation by angle $\varphi$ around axis $i$}. To obtain the direction of the spin axis in finite magnetic field $\bm{H}(\xi,\psi)$ we minimize the total energy
\begin{equation}\label{eq:Ftotal}
	\mathcal{F}_{total} = \mathcal{F}_a(\theta,\phi) + \mathcal{F}_Z (\theta,\phi,\xi,\psi) 
\end{equation}
with respect to $\theta$ and $\phi$. This approach assumes collinear two sublattice antiferromagnet and it allows us to obtain the direction of the spin axis as the direction of the minimal value of susceptibility tensor with respect to magnetic axes.\\
\indent In our simulations we construct the diagonal susceptibility tensor in such a way to reproduce the measured values $\chi_{a}$, $\chi_b$ and $\chi_c$ in tensor \eref{eq:tensorchiasbc} found in literature. Using $\chi_{m_1}=1.99\cdot 10^{-3}$~emu/mol, $\chi_{m_2}=2.0\cdot10^{-4}$~emu/mol and $\chi_{m_3}=2.39\cdot10^{-3}$~emu/mol gives $\chi_{a^*}=2.16\cdot10^{-3}$~emu/mol, $\chi_b=2.0\cdot10^{-4}$~emu/mol, $\chi_c=2.22\cdot10^{-3}$~emu/mol and $\chi_{ac}=\mp 2\cdot10^{-4}$~emu/mol, where $-$ and $+$ are signs for twins \textrm{I} and \textrm{II}, respectively. We then search for correct values of anisotropy constants $K_1$ and $K'_1$ by calculating the dependence of magnetization on magnetic field and by finding values of $K_1$ and $K'_1$ which reproduce the value of the spin flop field $H_{SF} \approx1.25$~T reported in literature from magnetization \cite{Heinrich-2003,Wheeler-2007, Rebello-2013} and ESR measurements \cite{Heinrich-2003}. Using the following values for anisotropy constants $K_1=5\cdot 10^4$~erg/mol and $K'_1=2.21\cdot 10^5$~erg/mol gives the dependence of magnetization on magnetic field shown in \fref{fig9} where spin flop is observed for magnetic field of $H=1.25$~T applied along the $m_2=\hat{b}$ axis. These values of anisotropy constants are of the order of magnitude typical for uniaxial antiferromagnet with this magnitude of $H_{SF}$ \cite{Gafvert-1977}. In \fref{fig9} we also show calculated magnetization for magnetic field applied in the $a^*$, $c$, $m_1$ and $m_3$ direction. We observe that for $H_{||b}\geq H_{SF}$ magnetization obtains the value it has for $H || m_3$. Absolute value of our calculated magnetization compares very well with the measured values from Ref. \onlinecite{Heinrich-2003}.\\
\indent The orientation of the spin axis obtained from our calculations for $H_{||b}\geq H_{SF}$ is in the direction of $m_3$, which $ = (-\eta,0,\gamma)$ for twin \textrm{I}, where $\eta=\sqrt{1-\gamma^2}$ and $\gamma=0.7584$ expressed in $(a^*, b, c)$ coordinate system. This means that spins have components along both $a$ and $c$ axis. If we express the observed direction $m_3$ in the monoclinic $(a, b, c)$ system, we get for spin axis direction $(m_{3a}, m_{3b}, m_{3c})=(- 0.6519, 0, 0.7712)$, where $m_{3a}$, $m_{3b}$ and $m_{3c}$ is the projection of the $m_3$ axis on $a$, $b$ and $c$ axis, respectively. This gives $|m_{3a}/m_{3c}|\approx 0.85$, which represents the ratio of magnetic moment along $a$ and $c$ axes in our model. Obtained result compares rather well to the ratio of projections of magnetic moment $m_a/m_c\approx 0.81$ observed in elastic neutron diffraction experiment performed on single crystal in $H= 6$~T applied along the $b$ axis of CuSb$_2$O$_6$ \cite{Wheeler-2007}. The orientation of spin axis for twin \textrm{I} in zero field and for $H_{|| m2}>H_{SF}$ is shown in \fref{fig10}. Results for other twins are given in table \ref{tab1}. Our result for the direction of spin axis is of course a consequence of the choice of the shape of phenomenological magnetocrystalline energy \eref{eq:Faniso}. To further prove that this choice is correct, we continue to calculate the angular dependence of torque using the same anisotropy energy.\\
\begin{figure}[t]
\centering
\includegraphics[width=1\columnwidth]{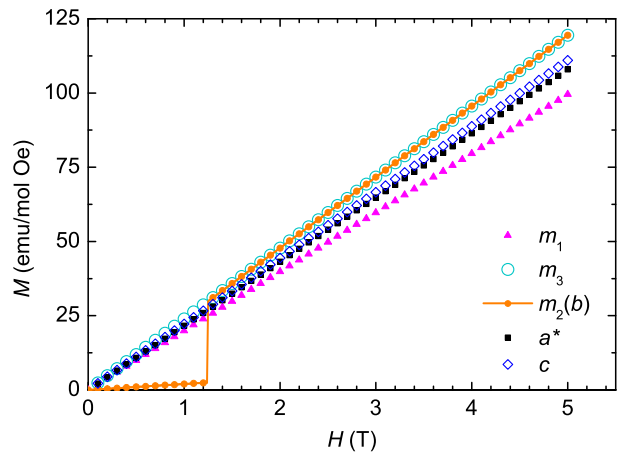}
\caption{Dependence of magnetization on magnetic field obtained from simulations for twin \textrm{I}.}
\label{fig9}
\end{figure}
\begin{figure}[t]
	\centering
		\includegraphics[width=1.00\columnwidth]{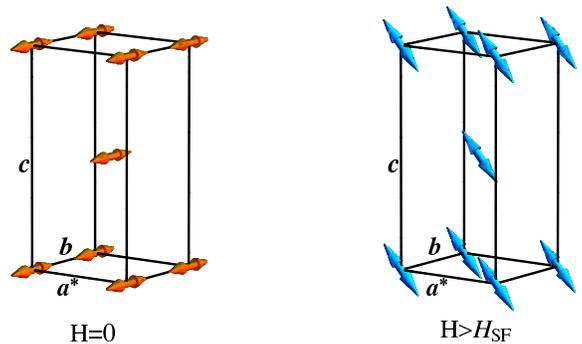}
	\caption{Orientation of spin axis in twin \textrm{I} in zero magnetic field and in $H>H_{SF}$ applied along easy axis direction obtained from our phenomenological treatment of magnetic anisotropy in CuSb$_2$O$_6$. Spin axis are represented as double-headed arrows since the direction of spins at specific sites and with respect to one another cannot be determined by this method. For other twins the direction of spin axes with respect to crystal axes is given in table \ref{tab1}.}
	\label{fig10}
\end{figure}
\indent Magnetization $\mathbf{M}=\bm{\hat{\chi}}\cdot \bm{H}$, expressed in emu Oe/mol is calculated for applied field $\bm{H}$ from the susceptibility tensor $\bm{\hat{\chi}}$ obtained from the minimization of total energy \eref{eq:Ftotal}, as decribed above. Magnetic torque is then calculated using expression $\boldsymbol{\tau}=m/(2M_{mol})\; \mathbf{M} \times \mathbf{H}$, where $m$ and $M_{mol}$ are mass and molar mass of the sample, respectively. This gives torque in units of erg or dyn~cm. We compare our measured torque with calculations for twin \textrm{I} only, since twins \textrm{I} and \textrm{II} are in great majority in our sample. Calculations for twin \textrm{II} for planes $a^*b$ and $bc$ in which the measurements were performed give the same torque curves as twin \textrm{I} due to symmetry, although the direction of spin axis in the flopped phase is different (see table \ref{tab1}). In Figs. \ref{fig6} and \ref{fig7} the results of calculations plotted by solid line are compared to the measured values. We multiply both measured and calculated torque by $(2 M_{mol})/(m H^2)$. The agreement of simulation and experiment is very good. Angular dependence given by expression \eqref{eq:torquePM} (dashed line in Figs. \ref{fig6} and \ref{fig7}), expected when no reorientation of spin axis takes place, describes the data well only in fields $H\lesssim 6$~kOe. Having assumed that axis $m_3$ is in the direction of the saddle point of anisotropy energy, we obtain the result in which the deformation of the angular dependence in higher fields is slightly larger in the $bc$ plane than in the $a^*b$ plane. Calculation with $m_1$ axis in the direction of the saddle point gives the opposite result which is not in agreement with the experiment. 
\section{Discussion}\label{sec:disc}
Our study of magnetic anisotropy of magnetically ordered state of CuSb$_2$O$_6$ was motivated by different reports on the direction of the easy axis in this system \cite{Nakua-1995, Kato-2002,Gibson-2004,Wheeler-2007,Rebello-2013}. Torque magnetometry is sensitive to magnetic susceptibility anisotropy, see \eqref{eq:torquePM}, and as such is a reliant experimental method for determining the direction of easy axis in simple collinear antiferromagnets \cite{Yosida-1951,Uozaki-2000}. We characterized our sample by performing ESR measurements to determine the presence and type of crystallographic twins which can have two different directions of the elongation in two crystallographically equivalent CuO$_6$ octahedra, as summarized in table \ref{tab1}. Twins of type \textrm{I} and \textrm{II} with $b$ axis as unique axis were shown to be dominant, and twins \textrm{III} and \textrm{IV} are found to make up less than $10\%$ of the sample. The dominance of twins \textrm{I} and \textrm{II} was also reported in literature \cite{Prokofiev-2003,Heinrich-2003,Prasai-2015}. From our torque magnetometry measurements we determined that in our sample, for which majority of twins are of type \textrm{I} and \textrm{II}, easy axis is along $b$ axis.\\
\indent Direction of easy axis in twins \textrm{III} and \textrm{IV} depends on weather these twins share common crystal $a$ and $b$ axes with twins \textrm{I} and \textrm{II}. There is a possibility that for twins \textrm{III} and \textrm{IV} $a$ axis is in the direction of the $b$ axis of twins \textrm{I} and \textrm{II}. Easy axis of twins \textrm{III} and \textrm{IV} are then also in the $b$ direction, but will be observed when field is applied in the direction of $a$ axis in twins \textrm{I} and \textrm{II}. This scenario trivially explains how easy axis behavior can be observed for field applied along both $a$ and $b$ axis. However, there is another explanation we put forward in our analysis which assumes that $a$ and $b$ axes are common for all twins, but unique crystallographic axes of monoclinic crystal structure are different. In twins \textrm{I} and \textrm{II} $b$ axis is unique axis of monoclinic crystal structure, and in twins \textrm{III} and \textrm{IV} it is $a$ axis. Twins \textrm{I} and \textrm{II} thus have $P 1 2_1/n 1$ symmetry while twins \textrm{III} and \textrm{IV} $P 2_1/n 1 1 $ symmetry. This conclusion is drawn from the analysis of $\mathbf{g}^2$ and $\bm{\hat{\chi}}$ tensors in relation to the Neumann's principle. We proceed with discussion assuming this scenario is realized in CuSb$_2$O$_6$.\\
\indent Magnetocrystalline anisotropy of twins \textrm{III} and \textrm{IV} must have the same shape as for twins \textrm{I} and \textrm{II} shown in \fref{fig8} when expressed in their respective macroscopic magnetic axes. Since it must be invariant to symmetry operations of the space group in the same way as magnetic susceptibility tensor, the $m_2$ axis in \fref{fig8} must correspond to the unique axes of monoclinic crystal structure, i.e. $m_2=\hat{a}$. $m_1$ and $m_3$ are then defined in the same way as for twins \textrm{I} and \textrm{II}, but are constrained to the $bc$ plane. Consequently, for twins \textrm{III} and \textrm{IV} easy axis is the $a$ axis and in $H_{||a}\geq H_{SF}$ the spins flop in the direction $m_3$, which is now in the $bc$ plane (see table \ref{tab1}). In our sample twins \textrm{III} and \textrm{IV} together amount to less than $10\%$ of the total sample, so we cannot quantitatively confirm this result. However, it would explain why in some cases reported in literature significant decrease of $\chi_a$ is observed along with the decrease of $\chi_b$. Directions of easy axis and spin axis in flopped phase for each twin obtained from our phenomenological model are summarized in table \ref{tab1}.\\
\indent For one sample of CuSb$_2$O$_6$ it was recently reported that in the AFM ordered state easy axis and spin flop was observed for both $a$ and $b$ directions \cite{Rebello-2013}. Indeed, the reported decrease of susceptibility along both $a$ and $b$ axis in that sample was comparable in magnitude corroborating this claim. Our analysis would explain this observation if different twins were present in this sample, regardless of the reason (common or rotated $a$ and 
$b$ axes) but authors claim that they managed to completely detwin their sample. In that case the anisotropy of their sample should be described by different type of anisotropy energy, such as found in tetragonal easy plane antiferromagnet Bi$_2$CuO$_4$ \cite{Herak-2010}. The symmetry of this anisotropy energy is higher than required for monoclinic CuSb$_2$O$_6$. In more recent papers by the same authors on, presumably, same samples the decrease of $\chi_a$ was significantly smaller than of $\chi_b$ \cite{Christian-2014} and some twinning of the samples was reported \cite{Prasai-2015}. The polishing of the surface which was used to detwin the samples in Ref. \onlinecite{Rebello-2013} and the method of taking the Laue images to determine the presence of twins was shown to be insufficient by Prokofiev et al. who also used this procedure and then went on to further investigate the apparently detwinned samples by X-ray counter diffractometer only to show that twinning still persists \cite{Prokofiev-2003}. In the same paper, Ref. \onlinecite{Prokofiev-2003}, different attempts to completely detwin the samples were reported to be unsuccessful and the conclusion was drawn that polishing only detwinns the surface. This adds to our claim that if two directions of easy axis, $a$ and $b$, appear in the sample of CuSb$_2$O$_6$, they must come from different crystallographic twins. 
In one single crystal neutron diffraction study, which determined the $b$ axis as easy axis, weak peaks from different twins were also observed, but these were, at least partially, removed from analysis.\\
\indent In order to unambiguously determine that the direction of easy axis can be different in different type of twin, as our analysis suggests, a systematic study of the contribution of different twins should be performed in samples where both $a$ and $b$ axis were observed as easy axis. However, such quantification of the contribution of different twins was not reported so far in literature, except in this study and in Ref. \onlinecite{Heinrich-2003}, where samples with dominant twins \textrm{I} and \textrm{II} are found, and where easy axis is the $b$ axis. A fully systematic study with different diffraction techniques, ESR and local magnetic techniques such as nuclear magnetic resonance, should also include the systematic pressure treatment of the samples to determine what is the factor that influences the contribution of different twins. While such analysis is beyond the scope of this work, it presents an interesting topic for future research.\\
\indent Finally, let us briefly comment on microscopic magnetic anisotropy that drives this system to specific arrangement of spins found in the ordered state. As stated in the introduction, for Cu$^{2+}$ spin $S=1/2$ systems anisotropy of the magnetically ordered states is driven by the weak anisotropic exchange. In CuSb$_2$O$_6$ antisymmetric anisotropic exchange i.e. Dzyaloshinskii-Moriya interaction \cite{Dzyaloshinsky-1958,*Moriya-1960,*Moriya-1960PR}, is not allowed between spins on chain and collinear arrangement of spins found in ground state suggests that it can be ignored. Symmetric anisotropic exchange then seems to be the reason behind the observed arrangement of spins. However, it was recently reported that spin-orbit coupling can induce weak single-ion anisotropy in spin $S=1/2$ transition metal ions and that this can be the driving mechanism behind arrangements of spins observed in some systems where standard approach fails \cite{Liu-2014}. Perhaps small spin flop field of 1.25~T observed for this sample is connected to the gap induced by this type of weak anisotropy.
\section{Conclusion}\label{sec:concl}
The antiferromagnetically ordered state of monoclinic CuSb$_2$O$_6$, which is still not satisfactorily understood, was studied using torque magnetometry as a sensitive probe of macroscopic magnetic anisotropy. Multiple crystallographic twins were detected by structural and ESR analysis which showed that in our sample dominant twins have $P 1 2_1/n 1$ symmetry with $b$ axis as unique axis. Torque measurements reveal that $b$ axis is the easy axis in this sample. However, our analysis of crystal and magnetic symmetry of different twins in monoclinic CuSb$_2$O$_6$ shows that both $a$ and $b$ axes can emerge as easy axis in different twins. Combining torque results with phenomenological approach to magnetocrystalline anisotropy energy we predict that when magnetic field $H\geq H_{SF}$ is applied along easy axis the spins flop in the direction of the macroscopic magnetic axis along which the susceptibility and $\mathbf{g}^2$ tensor have maximal values and that this direction is different for different twins. Our result nicely demonstrates how torque magnetometry can be used to determine the values of the spin flop field and the symmetry of the antiferromagnetically ordered state in both $H<H_{SF}$ and $H>H_{SF}$ by performing measurements in fields which are significantly smaller than $H_{SF}$. Results presented in this work offer possibility to reconcile different reports on easy axis direction found in literature and present a challenge for future studies of the crystal symmetry and magnetically ordered state of this system.
\begin{acknowledgments}
M. H. is grateful to Ognjen Milat and Kre\v{s}imir Salamon from the Institute of Physics in Zagreb for helpful discussion regarding the crystal structure and its symmetry. This work has been supported by the resources of the Croatian Ministry of Science, Education and Sports under Grant No. 035-0352843-2846 and in part by the Croatian Science Foundation under the Research project No. 1108.
\end{acknowledgments}
%

\bibliographystyle{apsrev4-1}
\bibliography{HerakRef}

\begin{thebibliography}{30}%
\makeatletter
\providecommand \@ifxundefined [1]{%
 \@ifx{#1\undefined}
}%
\providecommand \@ifnum [1]{%
 \ifnum #1\expandafter \@firstoftwo
 \else \expandafter \@secondoftwo
 \fi
}%
\providecommand \@ifx [1]{%
 \ifx #1\expandafter \@firstoftwo
 \else \expandafter \@secondoftwo
 \fi
}%
\providecommand \natexlab [1]{#1}%
\providecommand \enquote  [1]{``#1''}%
\providecommand \bibnamefont  [1]{#1}%
\providecommand \bibfnamefont [1]{#1}%
\providecommand \citenamefont [1]{#1}%
\providecommand \href@noop [0]{\@secondoftwo}%
\providecommand \href [0]{\begingroup \@sanitize@url \@href}%
\providecommand \@href[1]{\@@startlink{#1}\@@href}%
\providecommand \@@href[1]{\endgroup#1\@@endlink}%
\providecommand \@sanitize@url [0]{\catcode `\\12\catcode `\$12\catcode
  `\&12\catcode `\#12\catcode `\^12\catcode `\_12\catcode `\%12\relax}%
\providecommand \@@startlink[1]{}%
\providecommand \@@endlink[0]{}%
\providecommand \url  [0]{\begingroup\@sanitize@url \@url }%
\providecommand \@url [1]{\endgroup\@href {#1}{\urlprefix }}%
\providecommand \urlprefix  [0]{URL }%
\providecommand \Eprint [0]{\href }%
\providecommand \doibase [0]{http://dx.doi.org/}%
\providecommand \selectlanguage [0]{\@gobble}%
\providecommand \bibinfo  [0]{\@secondoftwo}%
\providecommand \bibfield  [0]{\@secondoftwo}%
\providecommand \translation [1]{[#1]}%
\providecommand \BibitemOpen [0]{}%
\providecommand \bibitemStop [0]{}%
\providecommand \bibitemNoStop [0]{.\EOS\space}%
\providecommand \EOS [0]{\spacefactor3000\relax}%
\providecommand \BibitemShut  [1]{\csname bibitem#1\endcsname}%
\let\auto@bib@innerbib\@empty
\bibitem [{\citenamefont {Nakua}\ \emph {et~al.}(1991)\citenamefont {Nakua},
  \citenamefont {Yun}, \citenamefont {Reimers}, \citenamefont {Greedan},\ and\
  \citenamefont {Stager}}]{Nakua-1991}%
  \BibitemOpen
  \bibfield  {author} {\bibinfo {author} {\bibfnamefont {A.}~\bibnamefont
  {Nakua}}, \bibinfo {author} {\bibfnamefont {H.}~\bibnamefont {Yun}}, \bibinfo
  {author} {\bibfnamefont {J.~N.}\ \bibnamefont {Reimers}}, \bibinfo {author}
  {\bibfnamefont {J.~E.}\ \bibnamefont {Greedan}}, \ and\ \bibinfo {author}
  {\bibfnamefont {C.~V.}\ \bibnamefont {Stager}},\ }\href
  {http://www.sciencedirect.com/science/article/pii/002245969190062M}
  {\bibfield  {journal} {\bibinfo  {journal} {J. Solid State Chem.}\ }\textbf
  {\bibinfo {volume} {91}},\ \bibinfo {pages} {105} (\bibinfo {year}
  {1991})}\BibitemShut {NoStop}%
\bibitem [{\citenamefont {Giere}\ \emph {et~al.}(1997)\citenamefont {Giere},
  \citenamefont {Brahimi}, \citenamefont {Deiseroth},\ and\ \citenamefont
  {Reinen}}]{Giere-1997}%
  \BibitemOpen
  \bibfield  {author} {\bibinfo {author} {\bibfnamefont {E.-O.}\ \bibnamefont
  {Giere}}, \bibinfo {author} {\bibfnamefont {A.}~\bibnamefont {Brahimi}},
  \bibinfo {author} {\bibfnamefont {H.~J.}\ \bibnamefont {Deiseroth}}, \ and\
  \bibinfo {author} {\bibfnamefont {D.}~\bibnamefont {Reinen}},\ }\href
  {http://www.sciencedirect.com/science/article/pii/S0022459697973746}
  {\bibfield  {journal} {\bibinfo  {journal} {J. Solid State Chem.}\ }\textbf
  {\bibinfo {volume} {131}},\ \bibinfo {pages} {263} (\bibinfo {year}
  {1997})}\BibitemShut {NoStop}%
\bibitem [{\citenamefont {Prokofiev}\ \emph {et~al.}(2003)\citenamefont
  {Prokofiev}, \citenamefont {Ritter}, \citenamefont {Assmus}, \citenamefont
  {Gibson},\ and\ \citenamefont {Kremer}}]{Prokofiev-2003}%
  \BibitemOpen
  \bibfield  {author} {\bibinfo {author} {\bibfnamefont {A.~V.}\ \bibnamefont
  {Prokofiev}}, \bibinfo {author} {\bibfnamefont {F.}~\bibnamefont {Ritter}},
  \bibinfo {author} {\bibfnamefont {W.}~\bibnamefont {Assmus}}, \bibinfo
  {author} {\bibfnamefont {B.~J.}\ \bibnamefont {Gibson}}, \ and\ \bibinfo
  {author} {\bibfnamefont {R.~K.}\ \bibnamefont {Kremer}},\ }\href
  {http://www.sciencedirect.com/science/article/pii/S0022024802020626}
  {\bibfield  {journal} {\bibinfo  {journal} {J. Crystal Growth}\ }\textbf
  {\bibinfo {volume} {247}},\ \bibinfo {pages} {457} (\bibinfo {year}
  {2003})}\BibitemShut {NoStop}%
\bibitem [{\citenamefont {Heinrich}\ \emph {et~al.}(2003)\citenamefont
  {Heinrich}, \citenamefont {Krug~von Nidda}, \citenamefont {Krimmel},
  \citenamefont {Loidl}, \citenamefont {Eremina}, \citenamefont {Ineev},
  \citenamefont {Kochelaev}, \citenamefont {Prokofiev},\ and\ \citenamefont
  {Assmus}}]{Heinrich-2003}%
  \BibitemOpen
  \bibfield  {author} {\bibinfo {author} {\bibfnamefont {M.}~\bibnamefont
  {Heinrich}}, \bibinfo {author} {\bibfnamefont {H.-A.}\ \bibnamefont {Krug~von
  Nidda}}, \bibinfo {author} {\bibfnamefont {A.}~\bibnamefont {Krimmel}},
  \bibinfo {author} {\bibfnamefont {A.}~\bibnamefont {Loidl}}, \bibinfo
  {author} {\bibfnamefont {R.~M.}\ \bibnamefont {Eremina}}, \bibinfo {author}
  {\bibfnamefont {A.~D.}\ \bibnamefont {Ineev}}, \bibinfo {author}
  {\bibfnamefont {B.~I.}\ \bibnamefont {Kochelaev}}, \bibinfo {author}
  {\bibfnamefont {A.~V.}\ \bibnamefont {Prokofiev}}, \ and\ \bibinfo {author}
  {\bibfnamefont {W.}~\bibnamefont {Assmus}},\ }\href {\doibase
  10.1103/PhysRevB.67.224418} {\bibfield  {journal} {\bibinfo  {journal} {Phys.
  Rev. B}\ }\textbf {\bibinfo {volume} {67}},\ \bibinfo {pages} {224418}
  (\bibinfo {year} {2003})}\BibitemShut {NoStop}%
\bibitem [{\citenamefont {Nakua}\ and\ \citenamefont
  {Greedan}(1995)}]{Nakua-1995}%
  \BibitemOpen
  \bibfield  {author} {\bibinfo {author} {\bibfnamefont {A.~M.}\ \bibnamefont
  {Nakua}}\ and\ \bibinfo {author} {\bibfnamefont {J.~E.}\ \bibnamefont
  {Greedan}},\ }\href
  {http://www.sciencedirect.com/science/article/pii/S002245968571331X}
  {\bibfield  {journal} {\bibinfo  {journal} {J. Solid State Chem.}\ }\textbf
  {\bibinfo {volume} {118}},\ \bibinfo {pages} {199} (\bibinfo {year}
  {1995})}\BibitemShut {NoStop}%
\bibitem [{\citenamefont {Koo}\ and\ \citenamefont {Whangbo}(2001)}]{Koo-2001}%
  \BibitemOpen
  \bibfield  {author} {\bibinfo {author} {\bibfnamefont {H.-J.}\ \bibnamefont
  {Koo}}\ and\ \bibinfo {author} {\bibfnamefont {M.-H.}\ \bibnamefont
  {Whangbo}},\ }\href
  {http://www.sciencedirect.com/science/article/pii/S0022459600989692}
  {\bibfield  {journal} {\bibinfo  {journal} {J. Solid State Chem.}\ }\textbf
  {\bibinfo {volume} {156}},\ \bibinfo {pages} {110} (\bibinfo {year}
  {2001})}\BibitemShut {NoStop}%
\bibitem [{\citenamefont {Kasinathan}\ \emph {et~al.}(2008)\citenamefont
  {Kasinathan}, \citenamefont {Koepernik},\ and\ \citenamefont
  {Rosner}}]{Kasinathan-2008}%
  \BibitemOpen
  \bibfield  {author} {\bibinfo {author} {\bibfnamefont {D.}~\bibnamefont
  {Kasinathan}}, \bibinfo {author} {\bibfnamefont {K.}~\bibnamefont
  {Koepernik}}, \ and\ \bibinfo {author} {\bibfnamefont {H.}~\bibnamefont
  {Rosner}},\ }\href {\doibase 10.1103/PhysRevLett.100.237202} {\bibfield
  {journal} {\bibinfo  {journal} {Phys. Rev. Lett.}\ }\textbf {\bibinfo
  {volume} {100}},\ \bibinfo {pages} {237202} (\bibinfo {year}
  {2008})}\BibitemShut {NoStop}%
\bibitem [{\citenamefont {Kato}\ \emph {et~al.}(2002)\citenamefont {Kato},
  \citenamefont {Kajimoto}, \citenamefont {Yoshimura}, \citenamefont {Kosuge},
  \citenamefont {Nishi},\ and\ \citenamefont {Kakurai}}]{Kato-2002}%
  \BibitemOpen
  \bibfield  {author} {\bibinfo {author} {\bibfnamefont {M.}~\bibnamefont
  {Kato}}, \bibinfo {author} {\bibfnamefont {K.}~\bibnamefont {Kajimoto}},
  \bibinfo {author} {\bibfnamefont {K.}~\bibnamefont {Yoshimura}}, \bibinfo
  {author} {\bibfnamefont {K.}~\bibnamefont {Kosuge}}, \bibinfo {author}
  {\bibfnamefont {M.}~\bibnamefont {Nishi}}, \ and\ \bibinfo {author}
  {\bibfnamefont {K.}~\bibnamefont {Kakurai}},\ }\href
  {http://journals.jps.jp/doi/abs/10.1143/JPSJS.71S.187} {\bibfield  {journal}
  {\bibinfo  {journal} {J. Phys. Soc. Jpn}\ }\textbf {\bibinfo {volume} {71}},\
  \bibinfo {pages} {187} (\bibinfo {year} {2002})}\BibitemShut {NoStop}%
\bibitem [{\citenamefont {Gibson}\ \emph {et~al.}(2004)\citenamefont {Gibson},
  \citenamefont {Kremer}, \citenamefont {Prokofiev}, \citenamefont {Assmus},\
  and\ \citenamefont {Ouladdiaf}}]{Gibson-2004}%
  \BibitemOpen
  \bibfield  {author} {\bibinfo {author} {\bibfnamefont {B.~J.}\ \bibnamefont
  {Gibson}}, \bibinfo {author} {\bibfnamefont {R.~K.}\ \bibnamefont {Kremer}},
  \bibinfo {author} {\bibfnamefont {A.~V.}\ \bibnamefont {Prokofiev}}, \bibinfo
  {author} {\bibfnamefont {W.}~\bibnamefont {Assmus}}, \ and\ \bibinfo {author}
  {\bibfnamefont {B.}~\bibnamefont {Ouladdiaf}},\ }\href
  {http://www.sciencedirect.com/science/article/pii/S0304885303013192}
  {\bibfield  {journal} {\bibinfo  {journal} {J. Magn. Magn. Matter.}\ }\textbf
  {\bibinfo {volume} {272-276}},\ \bibinfo {pages} {927} (\bibinfo {year}
  {2004})}\BibitemShut {NoStop}%
\bibitem [{\citenamefont {da~Silva~Wheeler}(2007)}]{Wheeler-2007}%
  \BibitemOpen
  \bibfield  {author} {\bibinfo {author} {\bibfnamefont {E.~M.}\ \bibnamefont
  {da~Silva~Wheeler}},\ }\href
  {http://xray.physics.ox.ac.uk/boothroyd/Theses/Elisa_Wheeler_thesis.pdf}
  {Ph.D. thesis},\ \bibinfo  {school} {Oxford University} (\bibinfo {year}
  {2007})\BibitemShut {NoStop}%
\bibitem [{\citenamefont {Torgashev}\ \emph {et~al.}(2003)\citenamefont
  {Torgashev}, \citenamefont {Shirokov}, \citenamefont {Prokhorov},
  \citenamefont {Gorshunov}, \citenamefont {Haas}, \citenamefont {Dressel},
  \citenamefont {Gibson}, \citenamefont {Kremer}, \citenamefont {Prokofiev},\
  and\ \citenamefont {Assmus}}]{Torgashev-2003}%
  \BibitemOpen
  \bibfield  {author} {\bibinfo {author} {\bibfnamefont {V.~I.}\ \bibnamefont
  {Torgashev}}, \bibinfo {author} {\bibfnamefont {V.~B.}\ \bibnamefont
  {Shirokov}}, \bibinfo {author} {\bibfnamefont {A.~S.}\ \bibnamefont
  {Prokhorov}}, \bibinfo {author} {\bibfnamefont {B.}~\bibnamefont
  {Gorshunov}}, \bibinfo {author} {\bibfnamefont {P.}~\bibnamefont {Haas}},
  \bibinfo {author} {\bibfnamefont {M.}~\bibnamefont {Dressel}}, \bibinfo
  {author} {\bibfnamefont {B.~J.}\ \bibnamefont {Gibson}}, \bibinfo {author}
  {\bibfnamefont {R.~K.}\ \bibnamefont {Kremer}}, \bibinfo {author}
  {\bibfnamefont {A.~V.}\ \bibnamefont {Prokofiev}}, \ and\ \bibinfo {author}
  {\bibfnamefont {W.}~\bibnamefont {Assmus}},\ }\href {\doibase
  10.1103/PhysRevB.67.134433} {\bibfield  {journal} {\bibinfo  {journal} {Phys.
  Rev. B}\ }\textbf {\bibinfo {volume} {67}},\ \bibinfo {pages} {134433}
  (\bibinfo {year} {2003})}\BibitemShut {NoStop}%
\bibitem [{\citenamefont {Rebello}\ \emph {et~al.}(2013)\citenamefont
  {Rebello}, \citenamefont {Smith}, \citenamefont {Neumeier}, \citenamefont
  {White},\ and\ \citenamefont {Yu}}]{Rebello-2013}%
  \BibitemOpen
  \bibfield  {author} {\bibinfo {author} {\bibfnamefont {A.}~\bibnamefont
  {Rebello}}, \bibinfo {author} {\bibfnamefont {M.~G.}\ \bibnamefont {Smith}},
  \bibinfo {author} {\bibfnamefont {J.~J.}\ \bibnamefont {Neumeier}}, \bibinfo
  {author} {\bibfnamefont {B.~D.}\ \bibnamefont {White}}, \ and\ \bibinfo
  {author} {\bibfnamefont {Y.-K.}\ \bibnamefont {Yu}},\ }\href {\doibase
  10.1103/PhysRevB.87.224427} {\bibfield  {journal} {\bibinfo  {journal} {Phys.
  Rev. B}\ }\textbf {\bibinfo {volume} {87}},\ \bibinfo {pages} {224427}
  (\bibinfo {year} {2013})}\BibitemShut {NoStop}%
\bibitem [{\citenamefont {Christian}\ \emph {et~al.}(2014)\citenamefont
  {Christian}, \citenamefont {Masunaga}, \citenamefont {Schye}, \citenamefont
  {Rebello}, \citenamefont {Neumeier},\ and\ \citenamefont
  {Yu}}]{Christian-2014}%
  \BibitemOpen
  \bibfield  {author} {\bibinfo {author} {\bibfnamefont {A.~B.}\ \bibnamefont
  {Christian}}, \bibinfo {author} {\bibfnamefont {S.~H.}\ \bibnamefont
  {Masunaga}}, \bibinfo {author} {\bibfnamefont {A.~T.}\ \bibnamefont {Schye}},
  \bibinfo {author} {\bibfnamefont {A.}~\bibnamefont {Rebello}}, \bibinfo
  {author} {\bibfnamefont {J.~J.}\ \bibnamefont {Neumeier}}, \ and\ \bibinfo
  {author} {\bibfnamefont {Y.-K.}\ \bibnamefont {Yu}},\ }\href {\doibase
  10.1103/PhysRevB.90.224423} {\bibfield  {journal} {\bibinfo  {journal} {Phys.
  Rev. B}\ }\textbf {\bibinfo {volume} {90}},\ \bibinfo {pages} {224423}
  (\bibinfo {year} {2014})}\BibitemShut {NoStop}%
\bibitem [{\citenamefont {Prasai}\ \emph {et~al.}(2015)\citenamefont {Prasai},
  \citenamefont {Rebello}, \citenamefont {Christian}, \citenamefont
  {Neumeier},\ and\ \citenamefont {Cohn}}]{Prasai-2015}%
  \BibitemOpen
  \bibfield  {author} {\bibinfo {author} {\bibfnamefont {N.}~\bibnamefont
  {Prasai}}, \bibinfo {author} {\bibfnamefont {A.}~\bibnamefont {Rebello}},
  \bibinfo {author} {\bibfnamefont {A.~B.}\ \bibnamefont {Christian}}, \bibinfo
  {author} {\bibfnamefont {J.~J.}\ \bibnamefont {Neumeier}}, \ and\ \bibinfo
  {author} {\bibfnamefont {J.~L.}\ \bibnamefont {Cohn}},\ }\href {\doibase
  10.1103/PhysRevB.91.054403} {\bibfield  {journal} {\bibinfo  {journal} {Phys.
  Rev. B}\ }\textbf {\bibinfo {volume} {91}},\ \bibinfo {pages} {054403}
  (\bibinfo {year} {2015})}\BibitemShut {NoStop}%
\bibitem [{\citenamefont {Bonner}\ and\ \citenamefont
  {Fisher}(1964)}]{Bonner-Fisher}%
  \BibitemOpen
  \bibfield  {author} {\bibinfo {author} {\bibfnamefont {J.~C.}\ \bibnamefont
  {Bonner}}\ and\ \bibinfo {author} {\bibfnamefont {M.~E.}\ \bibnamefont
  {Fisher}},\ }\href {\doibase 10.1103/PhysRev.135.A640} {\bibfield  {journal}
  {\bibinfo  {journal} {Phys. Rev.}\ }\textbf {\bibinfo {volume} {135}},\
  \bibinfo {pages} {A640} (\bibinfo {year} {1964})}\BibitemShut {NoStop}%
\bibitem [{\citenamefont {Johnston}\ \emph {et~al.}(2000)\citenamefont
  {Johnston}, \citenamefont {Kremer}, \citenamefont {Troyer}, \citenamefont
  {Wang}, \citenamefont {Kl\"umper}, \citenamefont {Bud'ko}, \citenamefont
  {Panchula},\ and\ \citenamefont {Canfield}}]{Johnston-2000}%
  \BibitemOpen
  \bibfield  {author} {\bibinfo {author} {\bibfnamefont {D.~C.}\ \bibnamefont
  {Johnston}}, \bibinfo {author} {\bibfnamefont {R.~K.}\ \bibnamefont
  {Kremer}}, \bibinfo {author} {\bibfnamefont {M.}~\bibnamefont {Troyer}},
  \bibinfo {author} {\bibfnamefont {X.}~\bibnamefont {Wang}}, \bibinfo {author}
  {\bibfnamefont {A.}~\bibnamefont {Kl\"umper}}, \bibinfo {author}
  {\bibfnamefont {S.~L.}\ \bibnamefont {Bud'ko}}, \bibinfo {author}
  {\bibfnamefont {A.~F.}\ \bibnamefont {Panchula}}, \ and\ \bibinfo {author}
  {\bibfnamefont {P.~C.}\ \bibnamefont {Canfield}},\ }\href {\doibase
  10.1103/PhysRevB.61.9558} {\bibfield  {journal} {\bibinfo  {journal} {Phys.
  Rev. B}\ }\textbf {\bibinfo {volume} {61}},\ \bibinfo {pages} {9558}
  (\bibinfo {year} {2000})}\BibitemShut {NoStop}%
\bibitem [{\citenamefont {Kittel}(2005)}]{Kittel}%
  \BibitemOpen
  \bibfield  {author} {\bibinfo {author} {\bibfnamefont {C.}~\bibnamefont
  {Kittel}},\ }\href@noop {} {\emph {\bibinfo {title} {Introduction to Solid
  State Physics}}},\ \bibinfo {edition} {8th}\ ed.\ (\bibinfo  {publisher}
  {John Wiley and Sons Inc.},\ \bibinfo {address} {New York},\ \bibinfo {year}
  {2005})\ p.\ \bibinfo {pages} {344}\BibitemShut {NoStop}%
\bibitem [{\citenamefont {Newnham}(2005)}]{Newnham}%
  \BibitemOpen
  \bibfield  {author} {\bibinfo {author} {\bibfnamefont {R.~E.}\ \bibnamefont
  {Newnham}},\ }\href@noop {} {\emph {\bibinfo {title} {Properties of Materials
  (Anisotropy, Symmetry, Structure)}}}\ (\bibinfo  {publisher} {Oxford
  University Press},\ \bibinfo {address} {New York, USA},\ \bibinfo {year}
  {2005})\BibitemShut {NoStop}%
\bibitem [{Note1()}]{Note1}%
  \BibitemOpen
  \bibinfo {note} {$2/m$ symmetry operations are rotation of 180 \protect
  \ensuremath {^\circ }around unique axis and mirror plane perpendicular to
  unique axis.}\BibitemShut {Stop}%
\bibitem [{\citenamefont {Yosida}(1951)}]{Yosida-1951}%
  \BibitemOpen
  \bibfield  {author} {\bibinfo {author} {\bibfnamefont {K.}~\bibnamefont
  {Yosida}},\ }\href {http://ptp.oxfordjournals.org/content/6/5/691.abstract}
  {\bibfield  {journal} {\bibinfo  {journal} {Prog. Theor. Phys.}\ }\textbf
  {\bibinfo {volume} {6}},\ \bibinfo {pages} {691} (\bibinfo {year}
  {1951})}\BibitemShut {NoStop}%
\bibitem [{\citenamefont {Uozaki}\ \emph {et~al.}(2000)\citenamefont {Uozaki},
  \citenamefont {Sasaki}, \citenamefont {Endo},\ and\ \citenamefont
  {Toyota}}]{Uozaki-2000}%
  \BibitemOpen
  \bibfield  {author} {\bibinfo {author} {\bibfnamefont {H.}~\bibnamefont
  {Uozaki}}, \bibinfo {author} {\bibfnamefont {T.}~\bibnamefont {Sasaki}},
  \bibinfo {author} {\bibfnamefont {S.}~\bibnamefont {Endo}}, \ and\ \bibinfo
  {author} {\bibfnamefont {N.}~\bibnamefont {Toyota}},\ }\href {\doibase
  10.1143/JPSJ.69.2759} {\bibfield  {journal} {\bibinfo  {journal} {Journal of
  the Physical Society of Japan}\ }\textbf {\bibinfo {volume} {69}},\ \bibinfo
  {pages} {2759} (\bibinfo {year} {2000})}\BibitemShut {NoStop}%
\bibitem [{\citenamefont {Herak}\ \emph {et~al.}(2010)\citenamefont {Herak},
  \citenamefont {Miljak}, \citenamefont {Dhalenne},\ and\ \citenamefont
  {Revcolevschi}}]{Herak-2010}%
  \BibitemOpen
  \bibfield  {author} {\bibinfo {author} {\bibfnamefont {M.}~\bibnamefont
  {Herak}}, \bibinfo {author} {\bibfnamefont {M.}~\bibnamefont {Miljak}},
  \bibinfo {author} {\bibfnamefont {G.}~\bibnamefont {Dhalenne}}, \ and\
  \bibinfo {author} {\bibfnamefont {A.}~\bibnamefont {Revcolevschi}},\ }\href
  {http://stacks.iop.org/0953-8984/22/i=2/a=026006} {\bibfield  {journal}
  {\bibinfo  {journal} {J. Phys.: Cond. Matter}\ }\textbf {\bibinfo {volume}
  {22}},\ \bibinfo {pages} {026006} (\bibinfo {year} {2010})}\BibitemShut
  {NoStop}%
\bibitem [{\citenamefont {Herak}(2011)}]{Herak-2011}%
  \BibitemOpen
  \bibfield  {author} {\bibinfo {author} {\bibfnamefont {M.}~\bibnamefont
  {Herak}},\ }\href
  {http://www.sciencedirect.com/science/article/pii/S0038109811003772}
  {\bibfield  {journal} {\bibinfo  {journal} {Solid State Comm.}\ }\textbf
  {\bibinfo {volume} {151}},\ \bibinfo {pages} {1588 } (\bibinfo {year}
  {2011})}\BibitemShut {NoStop}%
\bibitem [{\citenamefont {Skomski}(2008)}]{Skomski}%
  \BibitemOpen
  \bibfield  {author} {\bibinfo {author} {\bibfnamefont {R.}~\bibnamefont
  {Skomski}},\ }\href@noop {} {\emph {\bibinfo {title} {Simple Models of
  Magnetism}}}\ (\bibinfo  {publisher} {Oxford University Press},\ \bibinfo
  {address} {New York, USA},\ \bibinfo {year} {2008})\BibitemShut {NoStop}%
\bibitem [{Note2()}]{Note2}%
  \BibitemOpen
  \bibinfo {note} {$\protect \mathbf {R}_i(\varphi )$ is the rotation by angle
  $\varphi $ around axis $i$}\BibitemShut {NoStop}%
\bibitem [{\citenamefont {G\"{a}fvert}\ \emph {et~al.}(1977)\citenamefont
  {G\"{a}fvert}, \citenamefont {Lundgren}, \citenamefont {Westersrandh},\ and\
  \citenamefont {Beckman}}]{Gafvert-1977}%
  \BibitemOpen
  \bibfield  {author} {\bibinfo {author} {\bibfnamefont {U.}~\bibnamefont
  {G\"{a}fvert}}, \bibinfo {author} {\bibfnamefont {L.}~\bibnamefont
  {Lundgren}}, \bibinfo {author} {\bibfnamefont {B.}~\bibnamefont
  {Westersrandh}}, \ and\ \bibinfo {author} {\bibfnamefont {O.}~\bibnamefont
  {Beckman}},\ }\href
  {http://www.sciencedirect.com/science/article/pii/002236977790004X}
  {\bibfield  {journal} {\bibinfo  {journal} {J. Phys. Chem. Solids}\ }\textbf
  {\bibinfo {volume} {38}},\ \bibinfo {pages} {1333} (\bibinfo {year}
  {1977})}\BibitemShut {NoStop}%
\bibitem [{\citenamefont {Dzyaloshinsky}(1958)}]{Dzyaloshinsky-1958}%
  \BibitemOpen
  \bibfield  {author} {\bibinfo {author} {\bibfnamefont {I.}~\bibnamefont
  {Dzyaloshinsky}},\ }\href {\doibase
  http://dx.doi.org/10.1016/0022-3697(58)90076-3} {\bibfield  {journal}
  {\bibinfo  {journal} {Journal of Physics and Chemistry of Solids}\ }\textbf
  {\bibinfo {volume} {4}},\ \bibinfo {pages} {241 } (\bibinfo {year}
  {1958})}\BibitemShut {NoStop}%
\bibitem [{\citenamefont {Moriya}(1960{\natexlab{a}})}]{Moriya-1960}%
  \BibitemOpen
  \bibfield  {author} {\bibinfo {author} {\bibfnamefont {T.}~\bibnamefont
  {Moriya}},\ }\href {\doibase 10.1103/PhysRevLett.4.228} {\bibfield  {journal}
  {\bibinfo  {journal} {Phys. Rev. Lett.}\ }\textbf {\bibinfo {volume} {4}},\
  \bibinfo {pages} {228} (\bibinfo {year} {1960}{\natexlab{a}})}\BibitemShut
  {NoStop}%
\bibitem [{\citenamefont {Moriya}(1960{\natexlab{b}})}]{Moriya-1960PR}%
  \BibitemOpen
  \bibfield  {author} {\bibinfo {author} {\bibfnamefont {T.}~\bibnamefont
  {Moriya}},\ }\href {\doibase 10.1103/PhysRev.120.91} {\bibfield  {journal}
  {\bibinfo  {journal} {Phys. Rev.}\ }\textbf {\bibinfo {volume} {120}},\
  \bibinfo {pages} {91} (\bibinfo {year} {1960}{\natexlab{b}})}\BibitemShut
  {NoStop}%
\bibitem [{\citenamefont {Liu}\ \emph {et~al.}(2014)\citenamefont {Liu},
  \citenamefont {Koo}, \citenamefont {Xiang}, \citenamefont {Kremer},\ and\
  \citenamefont {Whangbo}}]{Liu-2014}%
  \BibitemOpen
  \bibfield  {author} {\bibinfo {author} {\bibfnamefont {J.}~\bibnamefont
  {Liu}}, \bibinfo {author} {\bibfnamefont {H.-J.}\ \bibnamefont {Koo}},
  \bibinfo {author} {\bibfnamefont {H.}~\bibnamefont {Xiang}}, \bibinfo
  {author} {\bibfnamefont {R.~K.}\ \bibnamefont {Kremer}}, \ and\ \bibinfo
  {author} {\bibfnamefont {M.-H.}\ \bibnamefont {Whangbo}},\ }\href {\doibase
  http://dx.doi.org/10.1063/1.4896148} {\bibfield  {journal} {\bibinfo
  {journal} {J. Chem. Phys.}\ }\textbf {\bibinfo {volume} {141}},\ \bibinfo
  {eid} {124113} (\bibinfo {year} {2014})}\BibitemShut {NoStop}%
\end{thebibliography}%

\end{document}